\documentclass[a4paper,11pt]{article}
\usepackage{a4wide,amsmath,amssymb}
\usepackage[english]{babel}
\date{}
\begin{document}
\begin{titlepage}
\title{
{\bf Covariant actions for models with non--linear twisted self--duality}
~\\
\medskip
\medskip
\medskip
\author{Paolo Pasti$^{a,b}$, Dmitri Sorokin$^{b,c}$ and Mario Tonin$^{a,b}$
~\\
~\\
{\small $^a$\it Dipartimento di Fisica e Astronomia ``Galileo Galilei", Universit\'a degli Studi di Padova}
~\\
\small{\it and}
~\\
{\small $^b$ \it INFN, Sezione di Padova, via F. Marzolo 8, 35131 Padova, Italia}
~\\
~\\
{\small $^c$ \it Department of Theoretical Physics, the University of the Basque Country UPV/EHU,}\\
{\small \it P.O. Box 644, 48080 Bilbao, Spain}
~\\
\small{\it and}
~\\
{\small IKERBASQUE, \it Basque Foundation for Science, Alameda Urquijo 36-5, 48011 Bilbao, Spain}
}
 }
\maketitle

\begin{abstract}
We describe  a  systematic  way of the generalization, to models with non--linear duality, of the space--time covariant and duality--invariant formulation of duality--symmetric theories in which the covariance of the action is ensured by the presence of a single auxiliary scalar field. It is shown that the duality--symmetric  action should be invariant  under the two  local symmetries characteristic of this approach, which impose constraints on the form of the action similar to those of Gaillard and Zumino and in the non--covariant formalism. We show that the (twisted) self--duality condition obtained from this action upon integrating its equations of motion can always be recast in a manifestly covariant form which is independent of the auxiliary scalar and thus corresponds to the conventional on--shell duality--symmetric covariant description of the same model. Supersymmetrization of this construction is briefly discussed.
\end{abstract}

\thispagestyle{empty}
\end{titlepage}

\section{Introduction}

Duality  invariance is an important symmetry that  arises in  many  models  of  physical  interest. A classical example is electrodynamics  without  sources  in  $D=4$ dimensions  where the $U(1)$ duality group mixes the field  strength  of the electric field with its Hodge  dual  identified with  the  field  strength of the (locally  defined) magnetic field. Another well known example is the $SL(2,R)$ duality symmetry of $D=10$ IIB supergravity.

Duality symmetries have been observed and the corresponding duality groups have been completely classified in $D=1,2,...,9$ supergravity models obtained by toroidal compactifications of $D=11$ supergravity.  An  important  case  is the toroidal compactification of $D=11$ supergravity on  seven  tori  that  leads  to  the celebrated D=4, N=8 supergravity with  $E_{7(7)}$  duality  group  \cite{deWit:1977fk,Cremmer:1979up,deWit:1982ig}.

Astonishingly,  explicit calculations \cite{Bern:2006kd,Bern:2007hh,Bern:2008pv,Bern:2009kf,Bern:2009kd,Dixon:2010gz} have proven that  $N=8$, $D=4$   supergravity  is  finite  at  the  perturbative  level  up  to  three and even four loops. These wonderful results have revived a great interest to  this theory in regards to an  old  question of  its  finiteness \cite{Deser:1977nt,Ferrara:1977mv,Deser:1978br,Kallosh:1980fi,Howe:1980th}. Since  supersymmetry  alone  is  not  sufficient  to  explains  these  results,  it has been natural to assume that the $ E_{7(7)}$ duality symmetry controls  remarkable  cancelations of divergent contributions to the supergravity amplitudes and,  perhaps, ensures  the possible finiteness of the theory \cite{Kallosh:2011dp,Kallosh:2011qt}. At  the perturbative  level,  this  symmetry  is a  global  continuous  symmetry, though it is broken to a discrete subgroup $E_{7(7)}(Z)$  by  non--perturbative  stringy  effects.

An  explanation  of  the  three--loop  finiteness  has  been suggested in \cite{Brodel:2009hu,Elvang:2010kc} by showing that the  only  possible supersymmetric candidate for the  counterterm   at  three  loops  violates  $E_{7(7)}$. The  same  argument  holds for  the candidate counterterms at five and six loops \cite{Bossard:2010bd,Beisert:2010jx} \footnote{At four loops there seem to be no supersymmetric counterterms.}.

The  arguments  in  \cite{Bossard:2010bd,Beisert:2010jx}, as well as in \cite{Berkovits:2006vc,Green:2006yu,Green:2010sp,Bossard:2011tq} suggest that  the first  divergent $E_{7(7)}$ invariant counterterm can  appear  at  seven loops.

Such a state of affairs leads one to assume that  if $N=8$, $D=4$ supergravity  is  perturbatively   finite,  the  reason  should  be  found  beyond  supersymmetry  and  $E_{7(7)}$ duality. However, in  our  opinion, before accepting this conclusion once and for all,  more  study  on  the  compatibility  of maximal  supersymmetry  with  $E_{7(7)}$ duality is needed. In other words, one should demonstrate whether the counterterms which may appear at higher loops are  consistent  from  this  point  of  view. Analogous issue has recently shown up in $N=4$, $D=4$ supergravity whose four--point amplitudes have been found to be free of divergencies at three loops \cite{Bern:2012cd,Tourkine:2012ip} in spite of the fact that supersymmetry admits at this order duality--invariant quantum  counterterms \cite{Bossard:2011tq}.

If the classical $N=4$ and $N=8$ supergravity do not allow for quantum deformations consistent with supersymmetry and duality invariance, then, as argued in \cite{Kallosh:2011dp,Kallosh:2011qt,Kallosh:2012ei}, this may be the reason of finiteness of these theories (at least at the corresponding loops).

In practice, one should understand i) how the possible counterterms (and their descendants) deform original linear duality relation between ``electric'' and  ``magnetic'' field strengths and ii) check whether this deformation is compatible with supersymmetry. The first problem has been addressed in \cite{Bossard:2011ij} and further developed in \cite{Carrasco:2011jv,Chemissany:2011yv}, where simpler examples of duality--invariant gauge theories with higher--order (Born--Infeld--like) and higher--derivative terms have been studied (see also \cite{Roiban:2012gi} for a related recent analysis at the quantum level). The second (supersymmetry) problem has been recently considered in \cite{Carrasco:2011jv,Broedel:2012gf} in the case of non--linear generalizations of $N=1,2$ $D=4$ supersymmetric Abelian gauge theories with $U(1)$ as the duality group, following earlier results of \cite{Cecotti:1986gb,Bagger:1996wp,Rocek:1997hi,Ketov:1998ku,Ketov:1998sx,Kuzenko:2000tg,Kuzenko:2000uh,Bellucci:2001hd,
Ketov:2001dq} based on the superfield formalism. To these results one should add the known examples of component non--linear duality--symmetric (Born--Infeld--type) Abelian gauge theories with 16 supersymmetries, namely, the $N=4$, $D=4$ supersymmetric Abelian Born--Infeld theory on the worldvolume of the $D3$--brane \cite{Cederwall:1996pv} and the $6d$ worldvolume theory of the M5--brane \cite{Howe:1996yn,Bandos:1997ui,Aganagic:1997zq,Howe:1997fb,Bandos:1997gm}.

When studying duality invariance of a theory one faces a well known problem that this symmetry usually directly manifests itself only on the mass shell, while the conventional Lagrangians are not invariant under the duality transformations. The reason is that only ``electric'' fields enter the Lagrangian, while their ``magnetic'' duals do not.  Instead, in order to guaranty the duality invariance of the field equations, the duality variation of the Lagrangian should satisfy a consistency requirement,  the  Gaillard--Zumino  condition \cite{Gaillard:1981rj}. This approach has been used in \cite{Bossard:2011ij} and \cite{Carrasco:2011jv,Chemissany:2011yv}.

To lift the duality invariance to the level of the action, the ``electric'' and ``magnetic'' fields should enter the Lagrangian on an equal footing, while the duality relation between them should arise on the mass shell as a consequence of equations of motion that follow from the Lagrangian. The latter guarantees that the number of the physical degrees of freedom remains intact.
One way to do this is to renounce the manifest space--time covariance of the action in favor of duality symmetry \cite{Zwanziger:1970hk,Deser:1976iy,Henneaux:1988gg,Schwarz:1993vs}. Note, however, that in such a formulation space--time (diffeomorphism or Lorentz) invariance is still  present  but is realized in a non--conventional way. Using this non--covariant formulation,  Hillmann \cite{Hillmann:2009zf} has obtained  the  duality  invariant action of $N=8$, $D=4$ supergravity, and Bossard, Hillmann and Nicolai \cite{Bossard:2010dq} have proved that the $E_{7(7)}$  symmetry is anomaly free in the perturbatively quantized theory.  In \cite{Bossard:2011ij} it has been suggested how one can reconstruct a non--linear duality--invariant action starting from a duality--invariant counterterm.

There is, however, a possibility of keeping manifest both, the duality and space--time symmetries in the action. This requires the introduction of auxiliary fields into the Lagrangian (see \cite{Pasti:2009xc} for a brief recent overview of different covariant formulations). The most economic way (dubbed the PST approach) is to introduce a single  auxiliary scalar field \cite{Pasti:1995tn}. In this formulation, in addition to the conventional gauge symmetry, the action is invariant under two extra local symmetries. One of them can be used to gauge away the auxiliary scalar and reduce the action to a non--manifestly Lorentz invariant form of the non--covariant approach. Another symmetry implies that some of the components of the gauge fields enter the action only under a total derivative and ensures the appearance of the duality relation as the general solution of the gauge field equations of motion.

The covariant PST approach unifies different non--covariant formulations \cite{Pasti:1996vs,Maznytsia:1998xw} and has proven to be extremely useful, in particular, for the construction of the action of the M5--brane in D=11 supergravity \cite{Bandos:1997ui}, which is an example of a non--linear self--dual (2,0) $6d$ gauge theory with 16 supersymmetries. So we hope that it may also be useful for making a further progress in pursuing the issue of the $E_{7(7)}$ and supersymmetry invariance of the $N=8$, $D=4$ supergravity effective action and the corresponding issue in less supersymmetric supergravities.

The  problem  is  to explicitly identify the possible  divergent  counterterms in $N=8$, $D=4$ supergravity and to show how to  write a consistent non--linear supersymmetric effective action, if any,  that  arises  from a  given counterterm and respects $E_{7(7)}$ duality symmetry.

As a preliminary study, the  purpose of  this  paper  is  to  solve  the problem considered  in \cite{Bossard:2011ij} in the framework of the covariant approach, namely, to  have  a  general  recipe  for constructing space--time covariant actions with  manifest duality symmetry at the non--linear level. Such a construction will include in the general framework the non--linear action for the $M5$--brane \cite{Bandos:1997ui,Aganagic:1997zq} and the corresponding on--shell covariant description of the $M5$--brane in the superembedding approach \cite{Howe:1996yn,Howe:1997fb}, as well as the manifestly duality--symmetric Lagrangian formulation of the Born--Infeld action for the D3--brane \cite{Berman:1997iz,Nurmagambetov:1998gp}. In this setting we will also clarify how the (twisted) self--duality condition obtained from the manifestly duality--symmetric action upon integrating its equations of motion can always be recast in a manifestly covariant form which is independent of the auxiliary scalar and thus corresponds to the conventional on--shell duality--symmetric covariant description of the same model.

This should set a stage for further analysis of the compatibility of supersymmetry with various possible non--linear deformations of a given duality symmetric theory, in particular, in the cases of extended supersymmetries and supergravities for which superfield methods are not applicable off the mass shell and/or have not yet been developed enough to include higher order corrections even on the mass shell.

The  paper  is  organized  as  follows.
In  Section 2, to  introduce  our  notation, we  review the covariant approach to theories with a linear self--duality condition. In Section 3 we  extend  the  approach to non--linear systems. This  is  done by starting  with a non--linear action  which is invariant, by  construction, under a local symmetry mentioned above, i.e. in such a way that some components of the gauge fields enter the action under a total derivative only.  The action is constructed as a series of local field functionals $I^{(k)}$, $k$ = 0, 1, ..., where $I^{(0)}$ is the term in the action which is quadratic  in the field strengths.  Then  one  imposes  the  condition that the  action  is invariant  also  under  the local symmetry which ensures the auxiliary nature of the PST scalar $a(x)$. This imposes  a  constraint on the local  functional $ I =  \sum  I^{(k)}$ that,  given $I^{(1)}$ ,  allows one to apply an  iterative  procedure to  determine $I^{(k)}$. In  Section 4  we derive the relation between the twisted self--duality condition obtained from the action in terms of the functionals $I^{(k)}$, that contain the auxiliary scalar $a(x)$, and a manifestly--covariant non--linear twisted self--duality condition which only involves the gauge field strengths (and derivatives thereof) and no auxiliary scalar.
Section  5 contains our  conclusions  and  includes a  non  exhaustive  discussion  of  the  compatibility  between supersymmetry and duality.

\section{\large\bf PST formulation of a linear duality--symmetric theory in $D=4$}

Consider a system of $N$ Abelian vector fields in $D=4$  described  by the  1--forms  $A^{r}(x)$ $(r =  1, ...,N)$
with the field  strengths  $ F^{r}  =  d A^{r}$ .  Call  $A^{\bar r}$   their  magnetic duals with  field strengths
$F^{\bar r}=dA^{\bar r}:=-\bullet \frac{2\delta S}{\delta F^r}$, where $S$ is an action constructed of the ``electric" field strengths $ F^{r}$  and $\bullet$ is the  Hodge  map\footnote{We have denoted the Hodge map by $\bullet$ instead of the conventional $\ast$, since we shall use the latter for denoting the twisted--duality conjugation. In  our  conventions,  a p--form $\phi_{(p)}$ in  $D$ dimensions is defined, in the vielbein basis,  as
$$  \phi_ {(p)} = \frac{1}{p!} e^{a_{1}}...e^{a_{p}}\phi_{a_{p}...a_{1}} $$
and its Hodge dual  is
$$ \bullet \phi_{(p)} = {1 \over {(D-p)!}}  e^{a_{1}}...e^{a_{D - p}}\epsilon_{a_{D - p}...a_{1}}{}^{b_{1}...b_{p}} \phi_{b_{p}...b_{1}}, $$
so  that,  in D=4 space--time  with  Minkowski  signature, $ \bullet \bullet  =  - 1$ . The external differential $d$ acts on the differential forms from the left}. For instance in the case of the Maxwell action $S_0=-\int d^4x\frac{1}{4}F_{\mu\nu}F^{\mu\nu}$ we have
 \begin{equation}\label{mag}
 F^{\bar r}_{\mu\nu}=(\bullet F^{r})_{\mu\nu} = {1 \over {2 }} \epsilon_{\mu\nu\rho \sigma}F^{r\rho \sigma}\,,\qquad F^{r}_{\mu\nu}=-(\bullet F^{\bar r})_{\mu\nu}=-{1 \over {2}} \epsilon_{\mu\nu\rho \sigma}F^{\bar r\rho \sigma}\,.
 \end{equation}

Now let us define  $  A^{i}  \equiv (A^{r},A^{\bar r})$  and  $  F^{i}  \equiv (F^{r},F^{\bar r})$ , $(i = 1, ..... 2N)$.
The  duality  group $ G  \subset  Sp(2N,R)$ acts  linearly on  $A^{i}$ ( and $F^{i}$ ).  The  vector  fields  $A^{r}$
can be coupled  to  gravity  (or  supergravity)  and to a  set  of  scalars, $\phi$,  and  fermions, $\psi$.

 In  the  presence  of scalars and  fermions  the  definition  of  the  field  strengths $  F^{i}$ can  be  generalized  as follows
 \begin{eqnarray}
 F^{i} =  d A^{i}  +  C^{i}
 \label{1}
 \end{eqnarray}
 where $C[\phi,\psi]^{i}$  are  two--forms. In supersymmetric theories such a redefinition is useful since it allows one to make the field strengths transform covariantly under supersymmetry.

The  scalars  parametrize  the  coset
$ G/ H $, where  $H$  is the  maximal  compact  subgroup  of  G, and  the  fermions belong to some  representation  of  $H$.
The  scalars $\phi$ are  described  by  the "bridges" $ {\cal V}(\phi)_{i}^{q}$
 where  the  index  $q$ spans  a representation  of $H$ whose (real) dimension is equal to that of $G$ labeled by $i$.
 One  can  define ${\cal V}_{i q } :=  ({\cal V}_{i}^{q})^* $  and  its inverse $ {\cal V}^{i p} $ such that $ {\cal V}_{i q} {\cal V}^{i p}  =  \delta_{q}^{p}$. Then  the  scalars  allow  to  define  an invertible  metric  in  $G$ given  by
\begin{equation}\label{1bis}
 G_{ij}  =  {\cal V}_{i}^{q} {\cal V}_{j q} +  c.c.
\end{equation}
Since G is a  subgroup  of  Sp(2N,R) one  can  define a  matrix $  \Omega^{ij} = - \Omega^{ji} $  with the  only  non vanishing
elements given  by $\Omega^{r \bar r} = - \Omega^{\bar r r} =  \delta ^{r  \bar r} $ and in a similar  way one can  define
$\Omega_{i j}$  so  that  $\Omega^{ij}\Omega_{jk} =  - \delta ^{i}_{k}$.  $\Omega^{ij}$  and  $\Omega_{ij}$  can  be  used  to  rise  and  lower
the  indices i , j ...
 Then  one  can  define  the  complex  structure
 \begin{eqnarray}
  J^{i}_{j} = G^{i k}\Omega_{k j} = \Omega^{ik}G_{kj}
  \label{2}
  \end{eqnarray}
    such  that
 $ J^{i}{}_{k}J^{k}{}_{j} =  -  \delta^{i}_{j} $  and finally  one  defines  the  "star  operation"  as  $$ \ast = J^{i}_{j} \bullet  $$
 so  that  $$ \ast  \ast = 1 . $$

With the use of the ``star operation" the duality relations between the electric and magnetic fields, eq. \eqref{mag}, take the form of the linear {\it  twisted  self--duality condition} on the field strength of  $A^{i}=(A^r,A^{\bar r})$.
\begin{eqnarray}
F^{i} = (\ast F)^{i}.
\label{3}
\end{eqnarray}
Note  that,  acting  on ({\ref{3}})  with the  differential  $d$ one  gets  the  field  equations of the vector fields. In addition, the  constraint ({\ref{3}}) implies that only half of the fields $A^i$, e.g. $A^{r}$, are independent, which ensures the correct number of the degrees of freedom of the theory.

Since $A^i$ transform linearly under the duality symmetry $G$ the duality constraint ({\ref{3}}) and, hence, the equations of motion are duality invariant. However, the conventional action constructed with a half number of the fields $A^i$ is not duality invariant. Instead, the duality symmetry manifests itself through the Gaillard--Zumino condition \cite{Gaillard:1981rj}  which should be satisfied by the duality variation of the action.

For studying properties of the duality--symmetric theory it is  useful to  have  an  action  that yields ({\ref{3}}) as a (consequence of) field  equations. However, since eq. ({\ref{3}}) is of the first order in derivatives, while usually the bosonic field equations are of the second order, constructing the duality--symmetric action turns out to be not a straightforward procedure.

One possibility is to renounce the requirement of manifest Lorentz invariance by splitting the $D=4$ Lorentz--vector indices of the fields  $A^{i}_{\mu}$. There are several ways of splitting the components of the $D=4$ vector, namely $4=(4-n)+n$ (where $n=1,2,3$). Each splitting results in a different non--covariant duality--symmetric action that produces eq.  ({\ref{3}}) (see \cite{Ho:2008nn,Pasti:2009xc,Chen:2010jgb,Huang:2011tu} for more details). In the original construction of \cite{Deser:1976iy}, which is closely related to the Hamiltonian description of the theory, the time--component   $A^{i}_{0}$  of  the  vectors $A^{i}_{\mu}$  $  ( \mu = (0 , m))  $ gets separated from their spacial components and does not appear in the action. Though this breaks the manifest Lorentz invariance, the action does invariant under a modified space--time symmetry which reduces to the conventional Lorentz symmetry on the mass shell \cite{Henneaux:1988gg,Schwarz:1993vs,Pasti:1995tn}.

 A manifestly Lorentz--covariant formulation of the duality--symmetric action that yields ({\ref{3}})  as  a  field  equation can be constructed following the approach proposed in \cite{Pasti:1995ii,Pasti:1995tn}. In  this  approach,  in  addition  to  the  physical  fields $ A^{i}_\mu$, the action contains an  auxiliary scalar field $a(x)$.  It enters the action through
the one--form $ v(x)$
{\footnote {The  signature  of  our  metric is (1, -1, -1, -1) }}
 \begin{eqnarray}
 v  =  {{da} \over { \sqrt { \partial_{\mu} a \partial^{\mu}a}}}
 \label{4}
 \end{eqnarray}
so  that $ v_{\mu}v^{\mu} = 1$  and
\begin{equation}\label{viv}
 v i_v + i_v v = 1\,,
 \end{equation}
where  $i_v $  is the  contraction with the vector  $ v^{\mu}\partial_{\mu} $ acting from  the  left, i.e.
\begin{equation}\label{iv}
i_{v}\phi_{(p)} = i_{v}\left({1 \over {p!}}e^be^{a_{1}}...e^{a_{p-1}}\phi_{a_{p-1}...a_{1} b}\right)= {1 \over {(p-1)!}} e^{a_{1}}...e^{a_{p-1}}\phi_{a_{p-1}...a_{1} b}v^{b}
\end{equation}
Acting  on  any p-form $X^{i}_{(p)} $ that transforms  as  a  vector  of  the  group $G$,   one  has  the  identities{\footnote  {For  notational  simplicity, sometimes in  expressions  like  $  (\ast X)^{i} $ we shall  drop  the  parentheses and write
 $  \ast X^{i} $.} }
 \begin{eqnarray}
 i_{v} \ast  = \ast v,   \qquad  \qquad   v \ast  = \ast i_{v}
 \label{5}
 \end{eqnarray}
so  that
\begin{eqnarray}
v i_{v} \ast  =  \ast i_{v} v,   \qquad  \qquad  i_v v \ast  =  \ast v i_{v}
 \label{6}
 \end{eqnarray}
 Using  these  identities  one  can decompose $F^i$ as follows
 \begin{eqnarray}
F^{i} = (v i_{v} + i_{v}v ) F^{i} =  v i_{v}F^{i} + \ast v i_{v}\ast F^{i} =  v i_{v}(F - \ast F)^{i} +(1 + \ast) (v i_{v}\ast F)^{i}
\label{7}
 \end{eqnarray}
In what follows we shall also use the following formulas  for the  variations $\delta v$ and $\delta i_v$ (acting  on  a p--form) with respect to $\delta a $:
 \begin{eqnarray}
 \delta v =  {1 \over \sqrt { \partial_{\mu}a \partial^{\mu}a}}\,i_{v} v (d \delta a)
 \label{7bis}
 \end{eqnarray}
 \begin{eqnarray}
 \delta i_{v}  =  {1 \over \sqrt { \partial_{\mu}a \partial^{\mu}a}} \ast i_{v} v (d \delta a) \ast
 \label{7ter}
 \end{eqnarray}

The  covariant action  $S_{0}$ can be  written  in various equivalent  ways, e.g.
\begin{eqnarray}
S_{0}  = {1 \over 8}\int \Omega_{ij} [ F^{i} \ast F^{j}  - (v i_{v}(F^{i} - \ast F^{i})) \ast (v i_{v} (F^{j} - \ast F^{j}))],
 \label{8}
 \end{eqnarray}
where for  simplicity  we have considered, for  the  moment,  the case with $C^{i} =0$, see eq. (\ref{1}). The general  case  will  be  considered
later.

  Using (\ref{7}), as well  as the  identities  (\ref{5}) and (\ref{6}), equation (\ref{8})  can be rewritten  as
 \begin{eqnarray}
S_{0}  = {1 \over 4}\int \Omega_{ij} [v i_{v}F^{i}\ast ( v i_{v}\ast F^{j}) - (v i_{v}\ast F^{i})\ast (v i_{v}\ast F^{j})]
 \label{9}
 \end{eqnarray}
or
 \begin{eqnarray}
S_{0}  = {1 \over 4}\int \Omega_{ij} [v i_{v}(F^{i} - \ast F^{i}) F^{j}]\,.
 \label{10}
 \end{eqnarray}
Eq. (\ref{9}) can also  be rewritten  as
 \begin{eqnarray}
S_{0}  =  - {1 \over 4} \int d^4 x \sqrt{g} G_{ij}[ (i_{v}F^{i})^{\mu} ( i_{v}\ast F^{j})_{\mu} - ( i_{v}\ast F^{i})^{\mu} ( i_{v}\ast F^{j})_{\mu}],
 \label{9bis}
 \end{eqnarray}
 where $g_{\mu\nu}(x)$ is the metric of the $4D$ space--time which for generality we consider to be curved.

An important property of the  action  $S_{0}$  is that the  $i_v A^{i}$ component of the gauge field enters this action only under the total derivative. Indeed,  $ i_{ v} A^{i}$ enters only  the  term  $ \int \Omega_{ij}(v i_{v} F^{i}) F^{j} $ of (\ref{10}) and the corresponding contribution is the total derivative, which is assumed to vanish at infinity
 $$ \int \Omega_{ij}v i_{v} (d (vi_{v}A^{i}) F^{j})
 = - \int \Omega_{ij}da d({ 1 \over \sqrt{(\partial a)^2}} i_{v}A^{i} ) F^{j} =  - \int d\left(\Omega^{ij}v i_{v} A^{i}  F^{j}\right)=0.$$
 The  independence of  $S_{0}$ from $i_{v}A^i$  is analogous to the absence
of  the  components $ A^{i}_{0}$ in  the  action  of  the  non--covariant  formulation. It implies that the action is invariant under the following local transformations of the gauge fields
\begin{eqnarray}
 \delta_{I} A^{i} =  d a  \Phi^{i},  \qquad   \delta_{I}  a  =  0,
 \label{13}
 \end{eqnarray}
  where  $\Phi^{i}(x) $ are scalar gauge parameters.

 Another symmetry of the action ensures that the field $a(x)$ is a pure gauge. It acts only on $a(x)$ and the duality symmetric gauge fields and leaves invariant other fields of the theory (scalars, fermions, metric etc.)
 \begin{eqnarray}
 \delta_{II}  a  =  \varphi(x),\qquad \delta_{II} A^{i} =  - \frac{1}{\sqrt{(\partial a)^2}}i_{v}(F^{i} - \ast F^{i}) \varphi(x),
 \label{14}
 \end{eqnarray}
 where $\varphi(x)$  is  a  local  gauge  parameter.  It is important  to  note  that the  $\delta_{II}$ variation
 of  $A^{i}$  is  proportional  to  the  self--duality  constraint,  which as we shall  see  is  a  consequence
 of  the  field  equation (\ref{11}),  so  that  it  vanishes  on  shell. The consequence of this fact is that on the mass shell the theory becomes manifestly Lorentz covariant without any need of the auxiliary field.

 The field equations of $A^i$ are
 \begin{eqnarray}
 d [v i_{v}(F^{i} - \ast F^{i})] = 0
 \label{11}
 \end{eqnarray}
 and the equation of motion of $a(x)$ is
  \begin{eqnarray}
 d [\Omega_{ij}{ 1 \over \sqrt{(\partial a)^2}} v i_{v}(F^{i} - \ast F^{i})i_{v}(F^{j}  - \ast F^{j})]  =  0.
 \label{12}
 \end{eqnarray}
  It can be obtained using the equations (\ref{7bis}) and (\ref{7ter}).

 One can check that eq. \eqref{12} is identically satisfied if eq. \eqref{11} holds. This reflects the fact that $a(x)$ is the auxiliary field. The general  solution  of (\ref{11}) is
\begin{equation}\label{sol}
v i_{v}(F^{i} - \ast F^{i}) =  d ( da X^{i})) = - da  dX^{i}
  \end{equation}
   where  $X^{i}(x)$  are  arbitrary  functions \footnote{Strictly speaking this is true only locally. In topologically non--trivial backgrounds $v i_{v}(F^{i} - \ast F^{i})$ may be closed but not exact one--form.}.
   On the other  hand,  under  (finite!)  transformations of the  symmetry \eqref{13},
  \begin{equation}\label{sol1}
  \delta_I [v i_{v}(F^{i} - \ast F^{i})]  = - da d \Phi^{i}
  \end{equation}
  so  that  a transformation  with the  parameter  $ \Phi^i  = - X^{i}$ allows us to eliminate $X^{i}$ from the right hand side of \eqref{sol} and get
 \begin{equation}\label{sol11}
 v i_{v}( F^{i} - \ast F^{i})  =  0.
 \end{equation}
   Moreover, since $(1 - \ast) F^{i} $ is  anti--selfdual, this equation also  implies
   \begin{equation}\label{sol2}
   i_{v} v ( F^{i} -  \ast  F^{i}) = 0.
   \end{equation}
   In view of \eqref{viv}, the equations \eqref{sol11} and \eqref{sol2} are equivalent to the twisted  self--duality  constraint (\ref{3}).

We have thus shown that the twisted self--duality relation follows from the covariant action as the solution of its equations of motion. Using the local symmetry \eqref{14} we can gauge fix the auxiliary field $a(x)$ to be
\begin{equation}\label{vm}
a(x) =  n_{\mu} x^{\mu},\qquad v_\mu=\frac{n_\mu}{\sqrt{n_\nu n^\nu} }
\end{equation}
where $n_{\mu}$  is a constant vector\footnote{Note that, though the gauge $a(x)=0$ is not directly admissible, since the action (\ref{8})  contains $\sqrt{(\partial a)^2}$ in the denominator, on can nevertheless reach this gauge
by handle a singularity in
the action in such a way that the ratio
${\partial_\mu a\,\partial^\nu a}/{\partial_\rho a\partial^\rho a}$
remains finite. This can be achieved by first imposing the gauge
fixing condition $a(x)=\epsilon\,x^\mu\,n_\mu$ and then sending the  parameter $\epsilon$ to zero.}. Depending on whether this vector is time--like or space--like, one reduces the PST action to different non--covariant formulations. For instance, when $n_{\mu}=\delta_\mu^0$, one recovers the non--covariant formulation of \cite{Deser:1976iy,Henneaux:1988gg,Schwarz:1993vs}. The non--conventional off--shell space--time invariance of the latter is explained by the necessity to keep intact the gauge condition \eqref{vm} under the Lorentz transformations, which is achieved by adding to the Lorentz variation of the gauge field $\delta_L A_\mu$ the compensating gauge transformation \eqref{14}
\begin{equation}\label{comp}
\delta A=\delta_{L}A+ \frac{1}{\sqrt{n_\mu n^\mu}}i_{n}(F^{i} -\ast F^{i}) n_\mu L^\mu{}_\nu x^\nu\,
\end{equation}
where $L^\mu{}_\nu$ are the infinitesimal parameters of the Lorentz transformation. Note that on the mass shell, i.e. when the twisted self--duality condition \eqref{sol11} is satisfied, the variation \eqref{comp} becomes the conventional Lorentz transformation of the gauge field.

 Up to  now  we  have  considered only  the case where  $ C^{i} = 0$ in (\ref{1}). The  general  case  can  be  easily  recovered by adding
 in  the  r.h.s.  of (\ref{8}) (  and (\ref{9}), and (\ref{10}) ) the  Wess--Zumino  term  $$ - {1 \over 2}\int \Omega_{ij} d A^{i} C^{j} .$$

\section{\large\bf PST action  with non--linear  duality  in $D=4$}

In the  previous  Section we considered  the case  in which the  magnetic  field  strengths $  F^{\bar r}$  are related to the  electric  ones  $F^{r}$  by the linear Hodge duality, or  equivalently the field  strengths  $F^{i}=(F^{r},F^{\bar r})$  satisfy  the  linear self--duality  constraint (\ref{3}). This  is  the  case in which the conventional action $S_{0}[F^{r}]$ is
quadratic  in  $F^{r}$. If in addition to $S_0$ an  action $ S = S_{0}  +  \hat S$  contains  terms $\hat S$  of higher  order in $F^{r}$  and/or  derivatives of  $F^{r}$, the  relation between $F^{\bar r}$ and $F^{r}$, \emph{i.e.}
$$ \bullet F^{\bar r}_{\mu\nu} = \delta^{\bar r}_{r} {{\delta  S}\over {\delta (F^{r})^{\mu\nu}}}$$
becomes  non--linear  in  $F^{r}$ and/or contains  derivatives of  $F^{r}$.
In  this  case the linear self--duality  constraint (\ref{3}) is  replaced  by  a non--linear (deformed) twisted self--duality  condition that  in  general can  be  expressed  as follows
\begin{eqnarray}
F^{i} - \lambda\bigl({{\delta W[F]} \over {\delta F}}\bigr)^{i} =  \ast (F - \lambda{{\delta W[F]} \over {\delta F}})^{i},\qquad
\bigl({{\delta W[F]} \over {\delta F}}\bigr)^{i} \equiv G^{ij}{1 \over 2} dx^{\mu}dx^{\nu}{{\delta W[F]} \over {\delta (F^{j})^{\nu \mu}}}\,.
\label{14bis}
\end{eqnarray}
where  $  W[ F ] $ is a  local  functional  of  $ F^i$  and  their  derivatives (as  well  as of other  fields) which is invariant  under  the    transformations  of  the  duality  group  $G$ and  $\lambda$  is a parameter of dimension $l^2$ which plays the role of a coupling constant characterizing a non--linear deformation of the Maxwell--like theory for which $\lambda=0$.
The  functional $ W[F]$ is, in  general, a  series  in $\lambda$ and $F$ \cite{Carrasco:2011jv}
\begin{eqnarray}
   W[F] =  \sum_{0}^{\infty} \lambda^{k}W^{(k)}[F]
  \label{27bis}
  \end{eqnarray}
The order $k$ of $\lambda$ is associated  with the dimension of terms  in $W^{(k)}$ in such a way that $\lambda W$ has the dimension $l^{-4}$. Duality--invariant counterterms of a quantum theory are examples of sources of the non--linearly deformed self--duality condition. Simple counterterm deformations, considered in \cite{Bossard:2011ij} are
$$
W[F]\sim C^2(\partial F)^2\,, \qquad W[F]\sim (F)^4,
$$
where $C$ is the $4d$ Weyl tensor.

In  this  Section  we  would like  to  extend  the  PST  approach to  the generic non--linear case. As we have seen in the previous Section, the self--duality condition which is derived from the PST action contains the auxiliary field $a(x)$, see eq. \eqref{sol11}. We have than shown that this relation is equivalent to the conventional covariant twisted self--duality condition \eqref{3} which does not contain $a(x)$. In the non--linear case we shall encounter and solve a similar problem, namely in the next Section we will demonstrate how the covariant non--linear twisted self--duality condition \eqref{14bis} is related to the one which we will now derive from the non--linear PST action\footnote{An example of the $6d$ counterpart of the condition \eqref{14bis} is the non--linearly self--dual field strength on the worldvolume of the M5--brane in the superembedding formulation \cite{Howe:1996yn,Howe:1997fb}. In \cite{Howe:1997vn} it was shown that the covariant non--linear self--duality condition, which is a consequence of a superembedding constraint, is related to a self--duality condition which follows from the M5--brane action \cite{Perry:1996mk,Bandos:1997ui,Aganagic:1997zq,Bandos:1997gm}. The latter either contains the (derivatives of) the auxiliary field $a(x)$, or (upon its gauge fixing) is not manifestly invariant under diffeomorphism (or Lorentz) transformations.}.

In  the  linear  case one of the possible forms  of  the PST  action was given  in eq. (\ref{9}) ( or (\ref{9bis})). Let us rewrite it as follows
 \begin{eqnarray}
S_{0}  = - {1 \over 2} \int d^4 x \sqrt{g} [G_{ij}{1 \over 2}  (i_{v}F^{i})^{\mu} (i_{v}\ast F^{j})_{\mu}  - {\cal L}^{(0)}],
 \label{15}
 \end{eqnarray}
 where
 \begin{eqnarray}
 {\cal L}^{(0)} = {1 \over 2 }G_{ij} (i_{v} \ast F^{i})^{\mu} (i_{v}\ast F^{j})_{\mu}.
\label{16}
\end{eqnarray}

As was shown in the  previous Section, the action \eqref{15} is invariant under the two local  symmetries \eqref{13} and \eqref{14}. This suggests to  consider in the  non--linear case the action
  \begin{eqnarray}
S  = - {1 \over 2} \int d^4 x \sqrt{g} [G_{ij}{1 \over 2}  (i_{v}F^{i})^{\mu}(i_{v}\ast F^{j})_{\mu}  - {\cal L}]
 \label{18}
 \end{eqnarray}
where  now
\begin{eqnarray}
{\cal L} = \sum_{0}^{\infty} \lambda^{k} {\cal L}^{(k)},
\label{19}
\end{eqnarray}
 $ {\cal L}^{(k)}$ are  local  functions of  $i_{v}(\ast F)^{i}$ (and, possibly, of their derivatives and of the other fields of the theory), ${\cal L}^{(0)} $ is  defined  in (\ref{16}).
 We shall  also  denote $$  I^{(k)} =  \int d^4x {\cal L}^{(k)} $$  and  $$  I  = \sum_{0}^{\infty} \lambda^{k}
I^{(k)}. $$

Since  ${\cal L}$ depends  on  $\ast F^i$  only  through  their contraction with $v$, \emph{i.e.} $ i_{v}\ast F^{i}$,  by  construction the  action (\ref{18}) is invariant  under the  symmetry  \eqref{13}. We  should also find the conditions under which  this  action  is  invariant under a non--linear generalization of the symmetry \eqref{14}. To find the form of this symmetry let us look at the equations of motion  of the  vector fields $A^{i}(x)$ and the auxiliary field $a(x)$. The vector field equations are
\begin{eqnarray}
d  \left[v \left(( i_{v}F)^{i} -  ({{\delta I } \over {\delta(i_{v} \ast F)}})^{i}\right)\right] =d  \left[v \left(( i_{v}(1  -  \ast )F)^{i} -  \lambda ({{\delta \hat I } \over {\delta(i_{v} \ast F)}})^{i}\right)\right] =  0,
\label{20}
\end{eqnarray}
where
\begin{equation}\label{hatI}
 \lambda \hat I = I  -  I^{(0)}  = \lambda \sum_{k=1}^{\infty}\lambda^{k-1} I^{(k)}
 \end{equation}
and $ {{\delta I^{(k)}} \over {\delta (i_{v} \ast F))}} $  are  the  one--forms
 \begin{eqnarray}
  ({{\delta I^{(k)} } \over {\delta(i_{v} \ast F)}})^{i}  = dx^{\mu} G^{ij} {{\delta I^{(k)}} \over {\delta(i_{v} \ast F)^{j \mu}}}.
\label{21}
\end{eqnarray}
Since $I^{(k)}$ (actually) depend on $v i_{v} \ast F^{i}$, one can write $ {{\delta I^{(k)}} \over {\delta (i_{v} \ast F))}} = i_{v} {{\delta I^{(k)}} \over {\delta (v i_{v} \ast F))}} $ and present
eq. (\ref{20}) in the form
\begin{eqnarray}
d  \left[v i_{v}(1  -  \ast )F^{i} -  \lambda v i_{v}({{\delta \hat I } \over {\delta(v i_{v} \ast F)}})^{i}\right] =  0,
\label{20bis}
\end{eqnarray}
where  $ {{\delta I^{(k)}} \over {\delta (v i_{v} \ast F))}}$  denote   the  two--forms defined  as  in (\ref{14bis}).

As in the linear case, eqs. \eqref{20} or \eqref{20bis} can be integrated and with the use of the local symmetry \eqref{13} result in the duality--like relations
\begin{equation}\label{20biss}
v \left(i_{v}F^{i} -  ({{\delta I } \over {\delta(i_{v} \ast F)}})^{i}\right)=v i_{v}\left((1  -  \ast )F^{i} -  \lambda ({{\delta \hat I } \over {\delta(v i_{v} \ast F)}})^{i}\right)=0\,.
\end{equation}

The  $a(x)$--field equation of motion is obtained from the action \eqref{18} using eqs. (\ref{7bis}) and (\ref{7ter}) and has the form
 \begin{eqnarray}
d \left\lbrace  {1 \over \sqrt{(\partial a)^2}}\Omega_{ij} v \left[ \left( (i_{v}\ast F^{i})(i_{v}\ast F^{j}) + (i_{v}F^{i})(i_{v}F^{j})\right) - 2 (i_{v}F^{i})
({{\delta I} \over {\delta (i_{v}\ast F)}})^{j}\right]\right\rbrace = 0.
\label{22}
\end{eqnarray}
Notice  that when  $ {\cal L}$ reduces  to  ${\cal L}^{(0)}$, at $\lambda = 0$, eqs. (\ref{20}) and (\ref{22}) reduce, respectively, to (\ref{11}) and (\ref{12}).

The form of the field equations (\ref{20}), \eqref{20biss} and (\ref{22}) prompts us that the non--linear generalization of the field variations under the second local symmetry (\ref{14}) should take the following form
\begin{eqnarray}
 \delta_{II} A^{i} =  - {1 \over \sqrt{(\partial a)^2}}[i_{v}F^{i} - ({{\delta I} \over {\delta(i_{v}\ast F)}})^{i} ]\,\varphi(x)  \qquad  ;  \qquad
  \delta_{II}  a  =  \varphi(x)
 \label{23}
 \end{eqnarray}
  The  variation  of  the  action under \eqref{23} is
  \begin{eqnarray}
  4\delta_{II} S = \int \delta_{II} a \,d \left\lbrace  {1 \over \sqrt{(\partial a)^2}}\Omega_{ij} v \left[ \left( (i_{v}\ast F^{i})(i_{v}\ast F^{j}) + (i_{v}F^{i})(i_{v}F^{j})\right)
  - 2 (i_{v}F^{i})({{\delta I} \over {\delta(i_{v}\ast F)}})^{j}\right]\right\rbrace
  \label{IInon}
 \end{eqnarray}
  \begin{eqnarray}
  +2 \int \Omega_{ij} \delta_{II}A^{i}\,d [ v ( i_{v}F^{j} - ( {{\delta I} \over {\delta(i_{v} \ast F)}})^{j})].
  \nonumber
  \end{eqnarray}
 For this variation to vanish, the following condition should hold
  \begin{eqnarray}
  d \left \lbrace {1 \over \sqrt{(\partial a)^2}}\Omega_{ij}\left [v\Big( (i_{v}\ast F^{i})(i_{v}\ast F^{j}) + (i_{v}F^{i})(i_{v}F^{j})\Big)
  - 2 v(i_{v}F^{i})({{\delta I} \over {\delta(i_{v}\ast F)}})^{j}\right]\right.
  \nonumber\\
  \left.-  v \left( i_{v}F^{i} -  ({{\delta I } \over {\delta( i_{v}\ast F)}})^{i}\right)
  \left(i_{v}F^{j} - ({{\delta I} \over {\delta(i_{v}\ast F)}}) ^{j} \right)\right\rbrace  =  0,
  \label{24}
  \end{eqnarray}
  which  can  be  simplified to
  \begin{eqnarray}
 d\left[\frac{1}{\sqrt{(\partial a)^2}} \,\Omega_{ij}  v \left((i_{v}\ast F^{i})(i_{v}\ast F^{j}) - ({{\delta I  } \over {\delta( i_{v}\ast F)}})^{i}
   ( {{\delta I} \over {\delta(i_{v}\ast F)}})^{j} \right)\right]  =  0.
  \label{25}
\end{eqnarray}
This equation is  the  fundamental  consistency  condition which is necessary  for the
action (\ref{18}) to be invariant under the local variations \eqref{23}. It ensures that $a(x)$ is a pure gauge degree of freedom. A similar condition has been found by Bossard and  Nicolai \cite{Bossard:2011ij}  in the non--covariant approach. The latter is obtained from \eqref{25} upon gauge fixing $a(x)=x^0$. This condition is  clearly  related  to the space--time invariance of the duality--symmetric construction and to the Gaillard--Zumino condition \cite{Gaillard:1981rj}.

 Eq.  (\ref{25}) is  automatically  satisfied  at  zero's order  in  $\lambda$.
At  first  order  in $\lambda$ one  has
\begin{eqnarray}
 d\left[\frac{1}{\sqrt{(\partial a)^2}}\,\Omega_{ij} v (i_{v}\ast F^{i}) ({{\delta I^{(1)} } \over {\delta( i_{v}\ast F)}})^{j}\right] =  0.
 \label{26}
 \end{eqnarray}
If  this  condition  is  satisfied by a certain choice of $I^{(1)}$,  the  consistency  condition (\ref{25}) imposes the constraint on the possible form of
$ {I}^{(2)}$ at  order $\lambda^2$
\begin{equation}\label{o2}
d\left[\frac{1}{\sqrt{(\partial a)^2}}\,\Omega_{ij} v \left(2(i_{v}\ast F^{i}) ({{\delta I^{(2)} } \over {\delta( i_{v}\ast F)}})^{j}+({{\delta I^{(1)} } \over {\delta( i_{v}\ast F)}})^{i}
   ( {{\delta I^{(1)}} \over {\delta(i_{v}\ast F)}})^{j} \right)\right] =  0,
\end{equation}
on $ {I}^{(3)}$ at order $\lambda^3$  and  so on. Solving these constraints one can reconstruct $I=\int{\cal L}$ order by order.

This iteration  procedure, however,  does  not  determine  $I=\int{\cal L}$  unambiguously. Indeed,  if at some  order $k$,  there  exists  an  action  $\bar I^{(k)} =
\int  \bar{\cal L}^{(k)}$ that  satisfies  eq. (\ref{26})  (with $I^{(1)}$   replaced  by  $ \bar I^{(k)}$), writing $  I^{(k)} +  c_{k} \bar I^{(k)}$ one  can  carry  on  the same procedure for $k'> k$ which will result in a  consistent  action that  now  depends on  the  arbitrary constant  $c_{k}$. This  arbitrariness  repeats  over  and  over  for  any  $\bar I^{k'}$  that satisfies  the  condition (\ref{26}).

Note that the invariance of the action under the gauge transformations \eqref{23} implies conditions on the form of the higher--order terms. Using the relations \eqref{6} one can rewrite eq. (\ref{20biss}) as follows
\begin{eqnarray}\label{biss}
v i_{v}(1  -  \ast )F^{i} =  \lambda v ({{\delta \hat I } \over {\delta(i_{v} \ast F)}})^{i}\quad\Rightarrow\nonumber\\
-i_{v}v (1  -  \ast )F^{i} =  \lambda\ast v ({{\delta \hat I } \over {\delta(i_{v} \ast F)}})^{i}\quad\Rightarrow\nonumber\\
(1  -  \ast )F^{i} =\lambda (1  -  \ast )v({{\delta \hat I } \over {\delta(i_{v} \ast F)}})^{i}=  \lambda (1  -  \ast )v i_{v}({{\delta \hat I } \over {\delta(v i_{v} \ast F)}})^{i}.
\end{eqnarray}
Since the left hand side of \eqref{biss} does not depend on $v$, also its right hand side should be $v$--independent, which imposes restrictions on the possible forms of $\hat I$. These restrictions are controlled by the local symmetry \eqref{23} and, hence, are a consequence of eq. \eqref{25}. Namely, the symmetry \eqref{23} can be used to gauge fix $v_\mu$ to be a constant vector as in \eqref{vm}. Then eq. \eqref{biss} implies that its right hand side must be Lorentz invariant on the mass shell, i.e. when the duality condition \eqref{20biss} is satisfied. This should be automatically so, since, as we have explained in the case of the linear self--duality, the on--shell Lorentz transformation \eqref{comp} of the gauge fields is the conventional one. If such, the right hand side of  \eqref{biss} must transform covariantly under the Lorentz symmetry and, therefore, can only be constructed of the Lorentz--covariant combinations of $F^i$ (and their derivatives).

This observation allows us to relate the higher--order terms in the action \eqref{18} to those of the  non--linear twisted self--duality condition (\ref{14bis}). Indeed, comparing eq. \eqref{biss} with (\ref{14bis}) we see that
\begin{equation}\label{WI}
(1  -  \ast )v{{\delta \hat I } \over {\delta(i_{v} \ast F)}}=(1  -  \ast )\frac{\delta W[F]} {\delta F}\,\quad {\rm or}\quad v{{\delta \hat I } \over {\delta(i_{v} \ast F)}}=vi_v(1  -  \ast )\frac{\delta W[F]} {\delta F}.
\end{equation}
Thus, knowing a higher--order deformation $W[F]$ of the original duality--symmetric theory, e.g. by quantum counterterms, one can obtain the form of the corresponding non--linear contributions to the duality--symmetric action and vice versa.

\section{\large\bf Relation between the two forms of the non--linear self--duality condition}

In  this  Section we shall demonstrate how to relate the self--duality  constraint  (\ref{14bis}) and the equation \eqref{20biss} obtained from the action (\ref{18}), i.e.  between   $W[F]$  and  $\hat I[F]$.

In  general, duality--invariant $W[F]$  depends  on  $F^{i}$  and  $\ast F^{i}$ or,  equivalently, on
$$F_{\pm }^{i} = {1 \over 2} (F  \pm \ast F)^{i}= {1 \over 2} (1  \pm \ast) F^{i}, $$
     so  that $  W[F] = W[ F^{+},F^{-}]$  and
\begin{equation} \label{Wpm}
      {{\delta W[F]} \over {\delta F}} = {1 \over 2} (1 -\ast) {{\delta W[F_{+},F_{-}]} \over {\delta F_{+}}} +  {1 \over 2} (1 + \ast) {{\delta W[F_{+},F_{-}]} \over {\delta F_{-}}}\,.
\end{equation}

   Substituting this  equation  into  (\ref{14bis}) we see  that
${{\delta W[F_{+},F_{-}]} \over {\delta F_{-}}}$ does not contribute, and the  self--duality  constraint (\ref{14bis}) becomes
\begin{eqnarray}
(1 - \ast)\Bigl( F^{i} - \lambda ({{\delta W[F_{+},F_{-}]} \over {\delta F_{+}}})^{i}\Bigr) = 0.
\label{27c}
\end{eqnarray}
Modulo different notation and approach, eq. (\ref{27c}) corresponds  to  eq. (4.2) of \cite{Carrasco:2011jv}.

Comparing \eqref{27c} with (\ref{WI}) we have
$$
(1  -  \ast) v ({{\delta \hat I } \over {\delta(i_{v} \ast F)}})^{i}=(1  -  \ast )\frac{\delta W[F]} {\delta F}=(1  -  \ast) ({{\delta W[F_{+},F_{-}]} \over {\delta F_{+}}})^{i}.
$$
To analyze the relation (\ref{WI}), let us introduce the identity (see eq. \eqref{7})
\begin{equation}\label{35}
F^{i} =   v i_{v}(F^{i}  -  \ast  F^{i})  - \lambda v({{\delta \hat I } \over {\delta(i_{v} \ast F)}})^{i} + ( 1  +  \ast )(v i_{v} \ast F^{i}) + \lambda v ({{\delta \hat I } \over {\delta(i_{v} \ast F)}})^{i}.
\end{equation}
Then on the mass shell \eqref{biss} we have
\begin{equation}\label{35+-}
F^{i} =  ( 1  +  \ast )(v i_{v} \ast F^{i}) + \lambda v ({{\delta \hat I } \over {\delta(i_{v} \ast F)}})^{i},
\end{equation}
\begin{equation}\label{35+}
F_{+}^{i} = (1  +  \ast )vi_{v}\ast F^{i} +  \frac{\lambda}{2}(1  +  \ast ) v ({{\delta \hat I } \over {\delta(i_{v} \ast F)}})^{i},
\end{equation}
and
$$
 F_{-}^{i} =  {\lambda\over 2}(1  - \ast) v ({{\delta \hat I } \over {\delta(i_{v} \ast F)}})^{i},
$$
which naturally coincides with (\ref{biss}). Equation \eqref{35+-}  tells us that, when the twisted self--duality relation holds, $F^i$  is a series in $vi_v\ast F^i$ and $\lambda$. Using this fact, one can carry out the following iteration procedure to reconstruct $\hat I=\sum_{k=1}^{\infty}\lambda^{k-1} I^{(k)}$ from a given counterterm $W[F]$ \eqref{27bis}. Possible  non vanishing  terms  $W^{(k)}$, $k\geq 1$,  are responsible for  the  arbitrariness in  $I$, pointed  out  at  the end  of  Section  2. Of course for consistency also these  $W^{(k)}$, on  shell  and  at  $\lambda = 0$, must satisfy  the  condition  (\ref{26}).

At the zero order in $\lambda$
$$
(1  -  \ast )\frac{\delta W[F]} {\delta F}|_{\lambda=0}=(1  -  \ast )f^{(0)}[vi_{v}\ast F^{i}],
$$
where $f^{(0)}$ is a known 2--form functional of $(1+\ast)vi_{v}\ast F^{i}$. This allows us, using \eqref{WI}, to reconstruct the first term $I^{(1)}$ of $\hat I$. Knowing $I^{(1)}$ we expand $\frac{\delta W[F]} {\delta F}$ to the first order in $\lambda$
\begin{equation}\label{first}
(1  -  \ast )\frac{\delta W[F]} {\delta F}=(1  -  \ast )\left(f^{(0)}[vi_{v}\ast F]+\lambda f^{(1)}[vi_{v}\ast F]+\lambda\frac{\delta W^{(1)}[F]} {\delta F}|_{\lambda=0}\right),
\end{equation}
where
$$f^{(1)}[vi_{v}\ast F]=\left[v {{\delta I^{(1)} } \over {\delta(i_{v} \ast F)}}\right]^{\mu\nu i}\frac{\delta^2W^{(0)}}{\delta [(1+\ast)vi_{v}\ast F]^{\mu\nu i}\delta F}|_{\lambda=0}$$
is a known 2--form functional of $vi_{v}\ast F$. Substituting eq. \eqref{first} into \eqref{WI} one reconstructs the second term $I^{(2)}$ of $\hat I$.

At the quadratic order in $\lambda$ the procedure for reconstructing $I^{(3)}$ becomes much more complicated since the expansion of $\frac{\delta W[F]} {\delta F}$ will have terms containing

$$[{{\delta I^{(1)} } \over {\delta(i_{v} \ast F)}}]^2,\quad{{\delta I^{(1)} } \over {\delta(i_{v} \ast F)}}\frac{\delta^2 W^{(1)}[F]} {\delta[(1+\ast)vi_{v} \ast F]\delta F}|_{\lambda=0},\quad{{\delta I^{(2)} } \over {\delta(i_{v} \ast F)}}\quad {\rm and}\quad
\frac{\delta W^{(2)}[F]} {\delta F}|_{\lambda=0}.$$
At the third and higher orders in $\lambda$ the complexity increases even more.

As a consistency check of the relations between $I^{(k)}$  and $W$, one should verify that the action functional $\hat I$ obtained in this way satisfies the  consistency  condition (\ref{25}) and whether this may impose additional restrictions on a possible form of $W$. Let us recall that, in the action \eqref{18} this condition insures that $a(x)$ is the completely auxiliary (pure gauge) field and that on the mass shell the self--duality condition can be brought to a space--time covariant form in terms of a duality--invariant functional $W[F]$ which does not depend on $a(x)$. To derive the constraint on $W[F]$ imposed by the consistency condition \eqref{25} note that on the mass shell \eqref{20biss} the latter takes the following form
\begin{eqnarray}\label{FF}
d\left[\frac{1}{\sqrt{(\partial a)^2}}\, \Omega_{ij}  v \left((i_{v}\ast F^{i})(i_{v}\ast F^{j}) - ((i_{v} F^{i})(i_{v} F^{j})\right)\right] \nonumber\\
 =d\left[\frac{1}{\sqrt{(\partial a)^2}}\, \Omega_{ij}  v i_{v}(1+\ast )F^{i})(i_{v}(1-\ast) F^{j}) \right]=  0,
\end{eqnarray}
which in turn, in view of \eqref{14bis} and \eqref{Wpm}, reduces to
\begin{equation}\label{F+}
\lambda d\left[ \frac{1}{\sqrt{(\partial a)^2}}\,\Omega_{ij} v \,(i_{v}F_+)^{i}\,\left(i_{v}\frac{\delta W}{\delta F_+}\right)^j \right]=  0,
\end{equation}

Though the statement that given any duality invariant $W[F]$ one can always reconstruct a corresponding duality--symmetric action looks plausible we have not found the generic proof that the constraint \eqref{F+} is satisfied by any choice of the duality invariant $W[F]$. We have checked the validity of \eqref{F+} for known examples of $W[F]$ which do not contain terms with derivatives of $F$. When $W[F]$ contains derivatives of $F$, the analysis becomes technically much more involved and we leave it for further study.

\section{Conclusion}

In this  paper we have described, in  a  systematic  way,  how to  extend  the  covariant and duality  invariant  PST  approach to models with  non--linear  duality.
It has been shown that  the duality--symmetric  action should be invariant  under  the two  local symmetries \eqref{13} and \eqref{23} characteristic  of  this  approach, which require  that  the  action is  given  by  eq. (\ref{18})  where  the local  functional $ I =\int \cal L$ depends  on  the fields  strengths $F^{i}$ only  through
$v i_{v}\ast F^{i}$ and  satisfies  the quadratic   constraint  (\ref{25}). This  constraint is related  to  the  Gaillard--Zumino  constraint
and, after a  suitable  gauge  fixing, coincides  with  the constraint found  in \cite{Bossard:2011ij}, in  the  framework  of  the non--covariant but  duality invariant  approach.

In the models with  non--linear duality, gauge fields are constrained  by  the {\it deformed twisted self--duality condition},  eq.
(\ref{14bis}). It  means  that  there  exists  a self--dual  two  form  $ h^{i} = F^{i}  -  ({{\delta W} \over {\delta F}})^{i}$ such  that
$  h^{i} = \ast h^{i} $,  where  $W[F,...]$ is a covariant and duality  invariant
local  functional  of $F^{i}$ (and  the  other  fields). As  a  further  result, in  this paper  we have  exploited  the  relation  between the functional  $W[F,...]$  and  the functional $ I $  that  constitutes the PST  action.
\if{}
All  the  functionals  $I^{(k)}$,  $k \geq 1 $,  are  determined from  $W[F]$  on shell.  In  particular  $W[F]$  determines  $I^{(1)}$ which  is  the  input  to start  the  iterative procedure described  at  the  end  of Section 3. Notice  that  the  fundamental  constraint (\ref{25}) implies  that  $I^{(1)}$ must  satisfy  the  condition  (\ref{26}).
Moreover  the  relation among $W[F]$  on shell and  the  $I^{(k)} $  implies  the  property  (\ref{27}) of  $I^{(k)}$  which  is  essential  to  assure that  the  field  equations of   the  theory yield  the  duality  constraint.
\fi

A  possible  application of the  approach  developed  in  this paper is the study of  the  consistent counterterms in  supersymmetric duality--invariant  models  and  in  particular in $N=8$, $D=4$ supergravity.
This  is  relevant  to  the issue  of  the  finiteness  of  this theory. The  question is whether $N=8$  supersymmetry  is  preserved upon a certain non--linear deformation of the classical theory. The authors of \cite{Bossard:2011ij} argued, on general grounds similar to those ensuring the diffeomorphism invariance and the absence of corresponding anomalies, that there might be no obstructions to find a deformed theory which is supersymmetric. To give more direct evidence for this argument, one should show that the Gaillard--Zumino or similar conditions, like eq. \eqref{25}, restricting the form of the action of the duality--symmetric theory are compatible with (deformed) supersymmetry transformations.

So far the compatibility of supersymmetry with non--linear self--duality has been explicitly demonstrated only for $N=1,2$ \cite{Cecotti:1986gb}--
\cite{Ketov:2001dq}, \cite{Carrasco:2011jv,Broedel:2012gf}  and $N=4$ \cite{Cederwall:1996pv} (D3--brane) Born--Infeld--like deformations of Abelian gauge theories with the duality group $U(1)$, and in the case of the M5--brane \cite{Bandos:1997ui,Aganagic:1997zq,Claus:1997cq} which is the non--linear (2,0) self--dual $6d$ gauge theory with 16 supersymmetries (i.e. $N=4$, from the $D=4$ perspective). However, supersymmetric examples of non--linear theories (including supergravities) with non--Abelian duality groups of the $E_7$ type have not been given yet. It should be mentioned that consistent couplings of \emph{external} supersymmetric Born--Infeld--like models to $N=1$ and 2 supergravities are known \cite{Kuzenko:2002vk,Kuzenko:2012ht}; however, an important issue which remains is whether non-linear deformations are possible for vector fields \emph{inside} supergravity multiplets, in particular, in $N=4,8$ supergravities.

At this point we would like to make a comment that non--linearities in field theories and, in particular, in supersymmetric ones are often associated with spontaneous symmetry and supersymmetry breaking. For instance, the Born--Infeld structure is a manifestation of partial supersymmetry breaking of a rigid extended supersymmetry \cite{Bagger:1996wp}. In this respect, Born--Infeld--like non--linearities in duality--symmetric effective action of $N=8$, $D=4$ supergravity, if appear, should have a different nature (e.g. stringy corrections), since there is no conventional field theories with more than 32 supersymmetries whose spontaneous breaking would result in a non--linear generalization of $N=8$, $D=4$ supergravity.
Restrictions on possible sources of the non--linear deformation of the twisted self--duality condition of $N=8$, $D=4$ supergravity imposed by supersymmetry and $E_{7(7)}$ are discussed in \cite{Kallosh:2012yy} \footnote{The authors are thankful to Renata Kallosh for sharing with them a draft of this paper.}.

There  are  two  complementary  approaches to  deal  with supersymmetric extensions of the duality--symmetric actions.
The  first is  the  standard  approach, in  which  the  action  depends only  on  the  ``electric"  fields.  In this  approach supersymmetry  is  manifest both  at  the  linear  level  and  in  the form of possible  candidate  counterterms,  but  the  duality  symmetry of the deformed action is  not  manifest and should be verified.  Given  a  supersymmetric  counterterm constructed of the  ``electric"  and ``magnetic" fields in a duality--invariant way, ref. \cite{Bossard:2011ij} has  described  an  iterative  procedure  further developed in \cite{Carrasco:2011jv} to  construct  a  non--linear
action for the  ``electric"  fields only, that satisfies the non--linear Gaillard--Zumino condition and, hence, retains the duality invariance. Since for the consistency with duality symmetry the non--linear deformation brings about an (infinite) series of new higher order terms, the supersymmetry of the whole construction should be rechecked.

Other approaches  deal  with  covariant  or  non--covariant  formulations  in  which  duality  symmetry  is  manifest. It  is clear  that  in  these  formulations supersymmetry  is  not  manifest  since the  number  of vector  fields is  doubled and only one  half  of  them  should  appear  in  the supersymmetry transformations  of  the  fermions.  Since  the  non--covariant  formulation  comes  from  a  gauge  fixing  of  the  covariant one, let  us discusse the supersymmetry issue in  the  framework  of  the  covariant  formulation.
In  models  with  linear  duality there  is  a  simple  recipe \cite{Schwarz:1993vs,Pasti:1995tn} how to  modify  the  supersymmetric  transformations of  the  fermions so  that  the PST action is invariant under this modified supersymmetry. In the supersymmetry variations of the fermions, the recipe prescribes to replace  the  field strengths $F^{i}$ with the following 2--form
\begin{eqnarray}
 K_{0}^{q} = [F^{i} - v i_{v}(F^{i}  - \ast F^{i})]V_i^q(\phi) =  (1  +  \ast) v i_{v} \ast F^{i}\,V_i^q(\phi),
 \label{360}
\end{eqnarray}
where $V_i^q(\phi)$ is the $G/H$ ``bridge" scalar field matrix determined in \eqref{1bis}.
Notice that $K_{0}^{i}$  is  self--dual, $ K_{0}^{q}  =  (\ast K_{0})^{q} $, and  that  on the shell of the linear duality constraint $ F^{i}  = \ast F^{i}$ the 2--form  $ K_{0}^{q}$ coincides with $F^{i}V_i^q(\phi)$. The property of  $K_{0}^{q}$ to be self--dual ensures that the supersymmetry transformations involve the right number of independent gauge fields.

For instance, in the simplest case of a $U(1)$--duality symmetric $N=1$ theory with no scalars ($V_i^q=\delta_i^q$) and one vector supermultiplet, duality--covariant $N=1$ supersymmetry variations look as follows
\begin{equation}\label{36}
\delta A^q_\mu=i\bar\psi\gamma_\mu\epsilon^q,\qquad \delta\psi=\frac{1}{8}K_0^{\mu\nu q}\gamma_{\mu\nu}\epsilon^q\,,\qquad q=1,2
\end{equation}
where $\psi(x)$ is the Majorana spinor and
\begin{equation}\label{susy}
\epsilon^q=i\varepsilon^{qs}\,\gamma_5\epsilon^s \qquad (\varepsilon^{12}=-\varepsilon^{21}=1)
\end{equation}
is the ``self--dual'' parameter of the rigid $N=1$, $D=4$ supersymmetry. It is easy to see that when the duality relation $F_{\mu\nu}^2=-\frac{1}{2}\varepsilon_{\mu\nu\rho\lambda}F^{1\rho\lambda}$ holds, the supersymmetry transformations \eqref{36} reduce to the conventional ones relating $A^1$ and $\psi$ with the Majorana spinor parameter $\epsilon^1$
$$
\delta A^1_\mu=i\bar\psi\gamma_\mu\epsilon^1,\qquad \delta\psi=\frac{1}{8}(F^{1\mu\nu }\gamma_{\mu\nu}\epsilon^1+F^{2\mu\nu}\gamma_{\mu\nu}\epsilon^2)=\frac{1}{4}F^{\mu\nu 1}\gamma_{\mu\nu}\epsilon^1.
$$
Let us also note that since the auxiliary field $a(x)$ does not have a superpartner, it should be invariant under the action of supersymmetry $\delta a(x)=0$. This, however, does not contradict the supersymmetry algebra, if one assumes that the translation of $a(x)$ produced by the commutator of two supersymmetry transformations acting on $a(x)$ is compensated by the local symmetry \eqref{14} \cite{Bandos:1997gd,Dall'Agata:1997db,Dall'Agata:1998va,DePol:2000re}
$$
(\delta_1\delta_2-\delta_2\delta_1) a(x)=\xi^\mu\partial_\mu a(x)-\varphi(x)=0\,.
$$

Now  the  problem  is how to  extend  the above prescription to  non--linear case. An  obvious  ansatz  would be to  replace $ K_{0}^{q}$  in (\ref{360})  with
\begin{eqnarray}
  K^{q} =  F^{q} -  v i_{v}\left[(1  - \ast) F ^{q} - \lambda (1  - \ast)\bigl({{ \delta W[F]} \over  {\delta F}}\bigr)^{q}\right] &=&
  (1  +  \ast) v i_{v}\left[ \ast F^{q}  +  {\lambda  \over 2} (1  -  \ast)\bigl({{ \delta W[F]} \over  {\delta F}}\bigr)^{q}\right]
 \nonumber\\
  &&+\frac{1}{2} \lambda(1  - \ast) \bigl({{ \delta W[F]} \over  {\delta F}}\bigr)^{q}\,.
\label{34}
\end{eqnarray}
Again,  on the duality shell \eqref{14bis}, $F^{q}  =  K^{q}$  but  now  $K^{q}$  is not  self--dual. However, the anti--self--dual part of $K^q$ does not enter the supersymmetry transformation \eqref{36} of the fermions, since its gamma--contraction with the self--dual supersymmetry parameter \eqref{susy} vanishes.

One may expect that this ansatz is incomplete and, in general, should also include terms of higher orders in fermionic fields. This is implicitly indicated by the analysis of rigid (2,0) supersymmetry transformations of the worldvolume fields of the kappa--symmetry gauge--fixed M5--brane carried out in \cite{Claus:1997cq}.
We hope to address the problem of supersymmetry in theories with non--linear duality in a future work.

\subsection*{Acknowledgements}
The authors are grateful to Renata Kallosh for having encouraged them to look at the generic problem of the construction of covariant actions with non--linear self--duality and for valuable discussions and comments. They also thank Igor Bandos for interest to this work and discussions. This work was partially supported by the INFN Special Initiative TV12 and by the Uni--PD Research Grant CPDA119349. D.S. is thankful to the Department of Theoretical Physics of the Basque Country University for hospitality and the IKERBASQUE Foundation for a visiting fellowship. D.S. also thanks the Organizers of the Program "The Mathematics and Applications of Branes in String and M--Theory" at the Isaac Newton Institute, Cambridge for invitation and hospitality during the final stage of this work.


\begin{thebibliography}{10}

\bibitem{deWit:1977fk}
B.~de~Wit and D.~Z. Freedman, ``{On SO(8) Extended Supergravity},''
\href{http://dx.doi.org/10.1016/0550-3213(77)90395-9}{{\em Nucl.Phys.}
  {\bfseries B130} (1977) 105}.

\bibitem{Cremmer:1979up}
E.~Cremmer and B.~Julia, ``{The SO(8) Supergravity},''
\href{http://dx.doi.org/10.1016/0550-3213(79)90331-6}{{\em Nucl.Phys.}
  {\bfseries B159} (1979) 141}.

\bibitem{deWit:1982ig}
B.~de~Wit and H.~Nicolai, ``{N=8 Supergravity},''
\href{http://dx.doi.org/10.1016/0550-3213(82)90120-1}{{\em Nucl.Phys.}
  {\bfseries B208} (1982) 323}.

\bibitem{Bern:2006kd}
Z.~Bern, L.~J. Dixon, and R.~Roiban, ``{Is N = 8 supergravity ultraviolet
  finite?},'' \href{http://dx.doi.org/10.1016/j.physletb.2006.11.030}{{\em
  Phys.Lett.} {\bfseries B644} (2007) 265--271},
  \href{http://arxiv.org/abs/hep-th/0611086}{{\ttfamily arXiv:hep-th/0611086
  [hep-th]}}.
7 pages, 5 figures, revtex.

\bibitem{Bern:2007hh}
Z.~Bern, J.~Carrasco, L.~J. Dixon, H.~Johansson, D.~Kosower, {\em et al.},
  ``{Three-Loop Superfiniteness of N=8 Supergravity},''
  \href{http://dx.doi.org/10.1103/PhysRevLett.98.161303}{{\em Phys.Rev.Lett.}
  {\bfseries 98} (2007) 161303},
\href{http://arxiv.org/abs/hep-th/0702112}{{\ttfamily arXiv:hep-th/0702112
  [hep-th]}}.

\bibitem{Bern:2008pv}
Z.~Bern, J.~Carrasco, L.~J. Dixon, H.~Johansson, and R.~Roiban, ``{Manifest
  Ultraviolet Behavior for the Three-Loop Four-Point Amplitude of N=8
  Supergravity},'' \href{http://dx.doi.org/10.1103/PhysRevD.78.105019}{{\em
  Phys.Rev.} {\bfseries D78} (2008) 105019},
\href{http://arxiv.org/abs/0808.4112}{{\ttfamily arXiv:0808.4112 [hep-th]}}.

\bibitem{Bern:2009kf}
Z.~Bern, J.~Carrasco, and H.~Johansson, ``{Progress on Ultraviolet Finiteness
  of Supergravity},''
\href{http://arxiv.org/abs/0902.3765}{{\ttfamily arXiv:0902.3765 [hep-th]}}.

\bibitem{Bern:2009kd}
Z.~Bern, J.~Carrasco, L.~J. Dixon, H.~Johansson, and R.~Roiban, ``{The
  Ultraviolet Behavior of N=8 Supergravity at Four Loops},''
  \href{http://dx.doi.org/10.1103/PhysRevLett.103.081301}{{\em Phys.Rev.Lett.}
  {\bfseries 103} (2009) 081301},
\href{http://arxiv.org/abs/0905.2326}{{\ttfamily arXiv:0905.2326 [hep-th]}}.

\bibitem{Dixon:2010gz}
L.~J. Dixon, ``{Ultraviolet Behavior of N = 8 Supergravity},''
\href{http://arxiv.org/abs/1005.2703}{{\ttfamily arXiv:1005.2703 [hep-th]}}.

\bibitem{Deser:1977nt}
S.~Deser, J.~Kay, and K.~Stelle, ``{Renormalizability Properties of
  Supergravity},''
\href{http://dx.doi.org/10.1103/PhysRevLett.38.527}{{\em Phys.Rev.Lett.}
  {\bfseries 38} (1977) 527}.

\bibitem{Ferrara:1977mv}
S.~Ferrara and B.~Zumino, ``{Structure of Conformal Supergravity},''
\href{http://dx.doi.org/10.1016/0550-3213(78)90548-5}{{\em Nucl.Phys.}
  {\bfseries B134} (1978) 301}.

\bibitem{Deser:1978br}
S.~Deser and J.~Kay, ``{Three loop counterterms for extended supergravity},''
\href{http://dx.doi.org/10.1016/0370-2693(78)90892-4}{{\em Phys.Lett.}
  {\bfseries B76} (1978) 400}.

\bibitem{Kallosh:1980fi}
R.~Kallosh, ``{Counterterms in extended supergravities},''
\href{http://dx.doi.org/10.1016/0370-2693(81)90964-3}{{\em Phys.Lett.}
  {\bfseries B99} (1981) 122--127}.

\bibitem{Howe:1980th}
P.~S. Howe and U.~Lindstrom, ``{Higher order invariants in extended
  supergravity},''
\href{http://dx.doi.org/10.1016/0550-3213(81)90537-X}{{\em Nucl.Phys.}
  {\bfseries B181} (1981) 487}.

\bibitem{Kallosh:2011dp}
R.~Kallosh, ``{$E_{7(7)}$ Symmetry and Finiteness of N=8 Supergravity},''
\href{http://arxiv.org/abs/1103.4115}{{\ttfamily arXiv:1103.4115 [hep-th]}}.

\bibitem{Kallosh:2011qt}
R.~Kallosh, ``{N=8 Counterterms and $E_{7(7)}$ Current Conservation},''
  \href{http://dx.doi.org/10.1007/JHEP06(2011)073}{{\em JHEP} {\bfseries 1106}
  (2011) 073},
\href{http://arxiv.org/abs/1104.5480}{{\ttfamily arXiv:1104.5480 [hep-th]}}.

\bibitem{Brodel:2009hu}
J.~Broedel and L.~J. Dixon, ``{R**4 counterterm and E(7)(7) symmetry in maximal
  supergravity},'' \href{http://dx.doi.org/10.1007/JHEP05(2010)003}{{\em JHEP}
  {\bfseries 1005} (2010) 003},
\href{http://arxiv.org/abs/0911.5704}{{\ttfamily arXiv:0911.5704 [hep-th]}}.

\bibitem{Elvang:2010kc}
H.~Elvang and M.~Kiermaier, ``{Stringy KLT relations, global symmetries, and
  $E_{7(7)}$ violation},''
  \href{http://dx.doi.org/10.1007/JHEP10(2010)108}{{\em JHEP} {\bfseries 1010}
  (2010) 108},
\href{http://arxiv.org/abs/1007.4813}{{\ttfamily arXiv:1007.4813 [hep-th]}}.

\bibitem{Bossard:2010bd}
G.~Bossard, P.~Howe, and K.~Stelle, ``{On duality symmetries of supergravity
  invariants},'' \href{http://dx.doi.org/10.1007/JHEP01(2011)020}{{\em JHEP}
  {\bfseries 1101} (2011) 020},
\href{http://arxiv.org/abs/1009.0743}{{\ttfamily arXiv:1009.0743 [hep-th]}}.

\bibitem{Beisert:2010jx}
N.~Beisert, H.~Elvang, D.~Z. Freedman, M.~Kiermaier, A.~Morales, {\em et al.},
  ``{E7(7) constraints on counterterms in N=8 supergravity},''
  \href{http://dx.doi.org/10.1016/j.physletb.2010.09.069}{{\em Phys.Lett.}
  {\bfseries B694} (2010) 265--271},
\href{http://arxiv.org/abs/1009.1643}{{\ttfamily arXiv:1009.1643 [hep-th]}}.

\bibitem{Berkovits:2006vc}
N.~Berkovits, ``{New higher-derivative $R^4$ theorems},''
  \href{http://dx.doi.org/10.1103/PhysRevLett.98.211601}{{\em Phys.Rev.Lett.}
  {\bfseries 98} (2007) 211601},
\href{http://arxiv.org/abs/hep-th/0609006}{{\ttfamily arXiv:hep-th/0609006
  [hep-th]}}.

\bibitem{Green:2006yu}
M.~B. Green, J.~G. Russo, and P.~Vanhove, ``{Ultraviolet properties of maximal
  supergravity},'' \href{http://dx.doi.org/10.1103/PhysRevLett.98.131602}{{\em
  Phys.Rev.Lett.} {\bfseries 98} (2007) 131602},
\href{http://arxiv.org/abs/hep-th/0611273}{{\ttfamily arXiv:hep-th/0611273
  [hep-th]}}.

\bibitem{Green:2010sp}
M.~B. Green, J.~G. Russo, and P.~Vanhove, ``{String theory dualities and
  supergravity divergences},''
  \href{http://dx.doi.org/10.1007/JHEP06(2010)075}{{\em JHEP} {\bfseries 1006}
  (2010) 075},
\href{http://arxiv.org/abs/1002.3805}{{\ttfamily arXiv:1002.3805 [hep-th]}}.

\bibitem{Bossard:2011tq}
G.~Bossard, P.~Howe, K.~Stelle, and P.~Vanhove, ``{The vanishing volume of D=4
  superspace},'' \href{http://dx.doi.org/10.1088/0264-9381/28/21/215005}{{\em
  Class.Quant.Grav.} {\bfseries 28} (2011) 215005},
\href{http://arxiv.org/abs/1105.6087}{{\ttfamily arXiv:1105.6087 [hep-th]}}.

\bibitem{Bern:2012cd}
Z.~Bern, S.~Davies, T.~Dennen, and Y.-t. Huang, ``{Absence of Three-Loop
  Four-Point Divergences in N=4 Supergravity},''
\href{http://arxiv.org/abs/1202.3423}{{\ttfamily arXiv:1202.3423 [hep-th]}}.

\bibitem{Tourkine:2012ip}
P.~Tourkine and P.~Vanhove, ``{An $R^4$ non-renormalisation theorem in $N=4$
  supergravity},''
\href{http://arxiv.org/abs/1202.3692}{{\ttfamily arXiv:1202.3692 [hep-th]}}.

\bibitem{Kallosh:2012ei}
R.~Kallosh, ``{On Absence of 3-loop Divergence in N=4 Supergravity},''
\href{http://arxiv.org/abs/1202.4690}{{\ttfamily arXiv:1202.4690 [hep-th]}}.

\bibitem{Bossard:2011ij}
G.~Bossard and H.~Nicolai, ``{Counterterms vs. Dualities},''
  \href{http://dx.doi.org/10.1007/JHEP08(2011)074}{{\em JHEP} {\bfseries 1108}
  (2011) 074},
\href{http://arxiv.org/abs/1105.1273}{{\ttfamily arXiv:1105.1273 [hep-th]}}.

\bibitem{Carrasco:2011jv}
J.~J.~M. Carrasco, R.~Kallosh, and R.~Roiban, ``{Covariant procedures for
  perturbative non-linear deformation of duality-invariant theories},''
  \href{http://dx.doi.org/10.1103/PhysRevD.85.025007}{{\em Phys.Rev.}
  {\bfseries D85} (2012) 025007},
\href{http://arxiv.org/abs/1108.4390}{{\ttfamily arXiv:1108.4390 [hep-th]}}.

\bibitem{Chemissany:2011yv}
W.~Chemissany, R.~Kallosh, and T.~Ortin, ``{Born-Infeld with Higher
  Derivatives},'' \href{http://dx.doi.org/10.1103/PhysRevD.85.046002}{{\em
  Phys.Rev.} {\bfseries D85} (2012) 046002},
\href{http://arxiv.org/abs/1112.0332}{{\ttfamily arXiv:1112.0332 [hep-th]}}.

\bibitem{Roiban:2012gi}
R.~Roiban and A.~Tseytlin, ``{On duality symmetry in perturbative quantum
  theory},''
\href{http://arxiv.org/abs/1205.0176}{{\ttfamily arXiv:1205.0176 [hep-th]}}.

\bibitem{Broedel:2012gf}
J.~Broedel, J.~J.~M. Carrasco, S.~Ferrara, R.~Kallosh, and R.~Roiban, ``{N=2
  Supersymmetry and U(1)-Duality},''
\href{http://arxiv.org/abs/1202.0014}{{\ttfamily arXiv:1202.0014 [hep-th]}}.

\bibitem{Cecotti:1986gb}
S.~Cecotti and S.~Ferrara, ``{Supersymmetric Born-Infeld Lagrangians},''
\href{http://dx.doi.org/10.1016/0370-2693(87)91105-1}{{\em Phys.Lett.}
  {\bfseries B187} (1987) 335}.

\bibitem{Bagger:1996wp}
J.~Bagger and A.~Galperin, ``{A New Goldstone multiplet for partially broken
  supersymmetry},'' \href{http://dx.doi.org/10.1103/PhysRevD.55.1091}{{\em
  Phys.Rev.} {\bfseries D55} (1997) 1091--1098},
\href{http://arxiv.org/abs/hep-th/9608177}{{\ttfamily arXiv:hep-th/9608177
  [hep-th]}}.

\bibitem{Rocek:1997hi}
M.~Rocek and A.~A. Tseytlin, ``{Partial breaking of global D = 4 supersymmetry,
  constrained superfields, and three-brane actions},''
  \href{http://dx.doi.org/10.1103/PhysRevD.59.106001}{{\em Phys.Rev.}
  {\bfseries D59} (1999) 106001},
\href{http://arxiv.org/abs/hep-th/9811232}{{\ttfamily arXiv:hep-th/9811232
  [hep-th]}}.

\bibitem{Ketov:1998ku}
S.~V. Ketov, ``{A Manifestly N=2 supersymmetric Born-Infeld action},''
  \href{http://dx.doi.org/10.1142/S0217732399000559}{{\em Mod.Phys.Lett.}
  {\bfseries A14} (1999) 501--510},
\href{http://arxiv.org/abs/hep-th/9809121}{{\ttfamily arXiv:hep-th/9809121
  [hep-th]}}.

\bibitem{Ketov:1998sx}
S.~V. Ketov, ``{Born-Infeld-Goldstone superfield actions for gauge fixed D-5
  branes and D-3 branes in 6-d},''
  \href{http://dx.doi.org/10.1016/S0550-3213(99)00239-4}{{\em Nucl.Phys.}
  {\bfseries B553} (1999) 250--282},
\href{http://arxiv.org/abs/hep-th/9812051}{{\ttfamily arXiv:hep-th/9812051
  [hep-th]}}.

\bibitem{Kuzenko:2000tg}
S.~M. Kuzenko and S.~Theisen, ``{Supersymmetric duality rotations},'' {\em
  JHEP} {\bfseries 0003} (2000) 034,
\href{http://arxiv.org/abs/hep-th/0001068}{{\ttfamily arXiv:hep-th/0001068
  [hep-th]}}.

\bibitem{Kuzenko:2000uh}
S.~M. Kuzenko and S.~Theisen, ``{Nonlinear selfduality and supersymmetry},''
  {\em Fortsch.Phys.} {\bfseries 49} (2001) 273--309,
\href{http://arxiv.org/abs/hep-th/0007231}{{\ttfamily arXiv:hep-th/0007231
  [hep-th]}}.

\bibitem{Bellucci:2001hd}
S.~Bellucci, E.~Ivanov, and S.~Krivonos, ``{Towards the complete N=2 superfield
  Born-Infeld action with partially broken N=4 supersymmetry},''
  \href{http://dx.doi.org/10.1103/PhysRevD.64.025014}{{\em Phys.Rev.}
  {\bfseries D64} (2001) 025014},
\href{http://arxiv.org/abs/hep-th/0101195}{{\ttfamily arXiv:hep-th/0101195
  [hep-th]}}.

\bibitem{Ketov:2001dq}
S.~V. Ketov, ``{Many faces of Born-Infeld theory},''
\href{http://arxiv.org/abs/hep-th/0108189}{{\ttfamily arXiv:hep-th/0108189
  [hep-th]}}.

\bibitem{Cederwall:1996pv}
M.~Cederwall, A.~von Gussich, B.~E. Nilsson, and A.~Westerberg, ``{The
  Dirichlet super three-brane in ten-dimensional type IIB supergravity},''
  \href{http://dx.doi.org/10.1016/S0550-3213(97)00071-0}{{\em Nucl.Phys.}
  {\bfseries B490} (1997) 163--178},
\href{http://arxiv.org/abs/hep-th/9610148}{{\ttfamily arXiv:hep-th/9610148
  [hep-th]}}.

\bibitem{Howe:1996yn}
P.~S. Howe and E.~Sezgin, ``{D = 11, p = 5},''
  \href{http://dx.doi.org/10.1016/S0370-2693(96)01672-3}{{\em Phys. Lett.}
  {\bfseries B394} (1997) 62--66},
\href{http://arxiv.org/abs/hep-th/9611008}{{\ttfamily arXiv:hep-th/9611008}}.

\bibitem{Bandos:1997ui}
I.~A. Bandos, K.~Lechner, A.~Y. Nurmagambetov, P.~Pasti, D.~P. Sorokin, and
  M.~Tonin, ``{Covariant action for the super-five-brane of M-theory},''
  \href{http://dx.doi.org/10.1103/PhysRevLett.78.4332}{{\em Phys. Rev. Lett.}
  {\bfseries 78} (1997) 4332--4334},
\href{http://arxiv.org/abs/hep-th/9701149}{{\ttfamily arXiv:hep-th/9701149}}.

\bibitem{Aganagic:1997zq}
M.~Aganagic, J.~Park, C.~Popescu, and J.~H. Schwarz, ``{World-volume action of
  the M-theory five-brane},''
  \href{http://dx.doi.org/10.1016/S0550-3213(97)00227-7}{{\em Nucl. Phys.}
  {\bfseries B496} (1997) 191--214},
\href{http://arxiv.org/abs/hep-th/9701166}{{\ttfamily arXiv:hep-th/9701166}}.

\bibitem{Howe:1997fb}
P.~S. Howe, E.~Sezgin, and P.~C. West, ``{Covariant field equations of the
  M-theory five-brane},''
  \href{http://dx.doi.org/10.1016/S0370-2693(97)00257-8}{{\em Phys. Lett.}
  {\bfseries B399} (1997) 49--59},
\href{http://arxiv.org/abs/hep-th/9702008}{{\ttfamily arXiv:hep-th/9702008}}.

\bibitem{Bandos:1997gm}
I.~A. Bandos, K.~Lechner, A.~Y. Nurmagambetov, P.~Pasti, D.~P. Sorokin, and
  M.~Tonin, ``{On the equivalence of different formulations of the M theory
  five-brane},'' \href{http://dx.doi.org/10.1016/S0370-2693(97)00784-3}{{\em
  Phys. Lett.} {\bfseries B408} (1997) 135--141},
\href{http://arxiv.org/abs/hep-th/9703127}{{\ttfamily arXiv:hep-th/9703127}}.

\bibitem{Gaillard:1981rj}
M.~K. Gaillard and B.~Zumino, ``{Duality Rotations for Interacting Fields},''
  \href{http://dx.doi.org/10.1016/0550-3213(81)90527-7}{{\em Nucl.Phys.}
  {\bfseries B193} (1981) 221}.
Dedicated to Andrei D. Sakharov on occasion of his 60th birthday.

\bibitem{Zwanziger:1970hk}
D.~Zwanziger, ``{Local Lagrangian quantum field theory of electric and magnetic
  charges},''
\href{http://dx.doi.org/10.1103/PhysRevD.3.880}{{\em Phys. Rev.} {\bfseries D3}
  (1971) 880}.

\bibitem{Deser:1976iy}
S.~Deser and C.~Teitelboim, ``{Duality Transformations of Abelian and
  Nonabelian Gauge Fields},''
\href{http://dx.doi.org/10.1103/PhysRevD.13.1592}{{\em Phys. Rev.} {\bfseries
  D13} (1976) 1592--1597}.

\bibitem{Henneaux:1988gg}
M.~Henneaux and C.~Teitelboim, ``{Dynamics of chiral (selfdual) P--forms},''
\href{http://dx.doi.org/10.1016/0370-2693(88)90712-5}{{\em Phys. Lett.}
  {\bfseries B206} (1988) 650}.

\bibitem{Schwarz:1993vs}
J.~H. Schwarz and A.~Sen, ``{Duality symmetric actions},''
  \href{http://dx.doi.org/10.1016/0550-3213(94)90053-1}{{\em Nucl. Phys.}
  {\bfseries B411} (1994) 35--63},
\href{http://arxiv.org/abs/hep-th/9304154}{{\ttfamily arXiv:hep-th/9304154}}.

\bibitem{Hillmann:2009zf}
C.~Hillmann, ``{E(7)(7) invariant Lagrangian of d=4 N=8 supergravity},''
  \href{http://dx.doi.org/10.1007/JHEP04(2010)010}{{\em JHEP} {\bfseries 1004}
  (2010) 010},
\href{http://arxiv.org/abs/0911.5225}{{\ttfamily arXiv:0911.5225 [hep-th]}}.

\bibitem{Bossard:2010dq}
G.~Bossard, C.~Hillmann, and H.~Nicolai, ``{E7(7) symmetry in perturbatively
  quantised N=8 supergravity},''
  \href{http://dx.doi.org/10.1007/JHEP12(2010)052}{{\em JHEP} {\bfseries 1012}
  (2010) 052},
\href{http://arxiv.org/abs/1007.5472}{{\ttfamily arXiv:1007.5472 [hep-th]}}.

\bibitem{Pasti:2009xc}
P.~Pasti, I.~Samsonov, D.~Sorokin, and M.~Tonin, ``{BLG-motivated Lagrangian
  formulation for the chiral two-form gauge field in D=6 and M5-branes},''
  \href{http://dx.doi.org/10.1103/PhysRevD.80.086008}{{\em Phys.Rev.}
  {\bfseries D80} (2009) 086008},
\href{http://arxiv.org/abs/0907.4596}{{\ttfamily arXiv:0907.4596 [hep-th]}}.

\bibitem{Pasti:1995tn}
P.~Pasti, D.~P. Sorokin, and M.~Tonin, ``{Duality symmetric actions with
  manifest space-time symmetries},''
  \href{http://dx.doi.org/10.1103/PhysRevD.52.R4277}{{\em Phys. Rev.}
  {\bfseries D52} (1995) 4277--4281},
\href{http://arxiv.org/abs/hep-th/9506109}{{\ttfamily arXiv:hep-th/9506109}}.

\bibitem{Pasti:1996vs}
P.~Pasti, D.~P. Sorokin, and M.~Tonin, ``{On Lorentz invariant actions for
  chiral p-forms},'' \href{http://dx.doi.org/10.1103/PhysRevD.55.6292}{{\em
  Phys. Rev.} {\bfseries D55} (1997) 6292--6298},
\href{http://arxiv.org/abs/hep-th/9611100}{{\ttfamily arXiv:hep-th/9611100}}.

\bibitem{Maznytsia:1998xw}
A.~Maznytsia, C.~R. Preitschopf, and D.~P. Sorokin, ``{Duality of selfdual
  actions},'' \href{http://dx.doi.org/10.1016/S0550-3213(98)00741-X}{{\em
  Nucl.Phys.} {\bfseries B539} (1999) 438--452},
\href{http://arxiv.org/abs/hep-th/9805110}{{\ttfamily arXiv:hep-th/9805110
  [hep-th]}}.

\bibitem{Berman:1997iz}
D.~Berman, ``{SL(2,Z) duality of Born-Infeld theory from nonlinear selfdual
  electrodynamics in six-dimensions},''
  \href{http://dx.doi.org/10.1016/S0370-2693(97)00919-2}{{\em Phys.Lett.}
  {\bfseries B409} (1997) 153--159},
\href{http://arxiv.org/abs/hep-th/9706208}{{\ttfamily arXiv:hep-th/9706208
  [hep-th]}}.

\bibitem{Nurmagambetov:1998gp}
A.~Nurmagambetov, ``{Duality symmetric three-brane and its coupling to type IIB
  supergravity},'' \href{http://dx.doi.org/10.1016/S0370-2693(98)00848-X}{{\em
  Phys.Lett.} {\bfseries B436} (1998) 289--297},
\href{http://arxiv.org/abs/hep-th/9804157}{{\ttfamily arXiv:hep-th/9804157
  [hep-th]}}.

\bibitem{Ho:2008nn}
P.-M. Ho and Y.~Matsuo, ``{M5 from M2},''
  \href{http://dx.doi.org/10.1088/1126-6708/2008/06/105}{{\em JHEP} {\bfseries
  0806} (2008) 105},
\href{http://arxiv.org/abs/0804.3629}{{\ttfamily arXiv:0804.3629 [hep-th]}}.

\bibitem{Chen:2010jgb}
W.-M. Chen and P.-M. Ho, ``{Lagrangian Formulations of Self-dual Gauge Theories
  in Diverse Dimensions},''
  \href{http://dx.doi.org/10.1016/j.nuclphysb.2010.04.015}{{\em Nucl.Phys.}
  {\bfseries B837} (2010) 1--21},
\href{http://arxiv.org/abs/1001.3608}{{\ttfamily arXiv:1001.3608 [hep-th]}}.

\bibitem{Huang:2011tu}
W.-H. Huang, ``{Lagrangian of Self-dual Gauge Fields in Various
  Formulations},''
\href{http://arxiv.org/abs/1111.5118}{{\ttfamily arXiv:1111.5118 [hep-th]}}.

\bibitem{Pasti:1995ii}
P.~Pasti, D.~P. Sorokin, and M.~Tonin, ``{Note on manifest Lorentz and general
  coordinate invariance in duality symmetric models},''
  \href{http://dx.doi.org/10.1016/0370-2693(95)00463-U}{{\em Phys. Lett.}
  {\bfseries B352} (1995) 59--63},
\href{http://arxiv.org/abs/hep-th/9503182}{{\ttfamily arXiv:hep-th/9503182}}.

\bibitem{Howe:1997vn}
P.~S. Howe, E.~Sezgin, and P.~C. West, ``{The six-dimensional self-dual
  tensor},'' \href{http://dx.doi.org/10.1016/S0370-2693(97)00365-1}{{\em Phys.
  Lett.} {\bfseries B400} (1997) 255--259},
\href{http://arxiv.org/abs/hep-th/9702111}{{\ttfamily arXiv:hep-th/9702111}}.

\bibitem{Perry:1996mk}
M.~Perry and J.~H. Schwarz, ``{Interacting chiral gauge fields in six
  dimensions and Born-Infeld theory},''
  \href{http://dx.doi.org/10.1016/S0550-3213(97)00040-0}{{\em Nucl. Phys.}
  {\bfseries B489} (1997) 47--64},
\href{http://arxiv.org/abs/hep-th/9611065}{{\ttfamily arXiv:hep-th/9611065}}.

\bibitem{Claus:1997cq}
P.~Claus, R.~Kallosh, and A.~Van~Proeyen, ``{M five-brane and superconformal
  (0,2) tensor multiplet in six-dimensions},''
  \href{http://dx.doi.org/10.1016/S0550-3213(98)00137-0}{{\em Nucl.Phys.}
  {\bfseries B518} (1998) 117--150},
\href{http://arxiv.org/abs/hep-th/9711161}{{\ttfamily arXiv:hep-th/9711161
  [hep-th]}}.

\bibitem{Kuzenko:2002vk}
S.~M. Kuzenko and S.~A. McCarthy, ``{Nonlinear selfduality and supergravity},''
  {\em JHEP} {\bfseries 0302} (2003) 038,
\href{http://arxiv.org/abs/hep-th/0212039}{{\ttfamily arXiv:hep-th/0212039
  [hep-th]}}.

\bibitem{Kuzenko:2012ht}
S.~M. Kuzenko, ``{Nonlinear self-duality in N = 2 supergravity},''
\href{http://arxiv.org/abs/1202.0126}{{\ttfamily arXiv:1202.0126 [hep-th]}}.

\bibitem{Kallosh:2012yy}
R.~Kallosh and T.~Ortin, ``{New E77 invariants and amplitudes},''
\href{http://arxiv.org/abs/1205.4437}{{\ttfamily arXiv:1205.4437 [hep-th]}}.

\bibitem{Bandos:1997gd}
I.~A. Bandos, N.~Berkovits, and D.~P. Sorokin, ``{Duality symmetric
  eleven-dimensional supergravity and its coupling to M-branes},''
  \href{http://dx.doi.org/10.1016/S0550-3213(98)00102-3}{{\em Nucl.Phys.}
  {\bfseries B522} (1998) 214--233},
\href{http://arxiv.org/abs/hep-th/9711055}{{\ttfamily arXiv:hep-th/9711055
  [hep-th]}}.

\bibitem{Dall'Agata:1997db}
G.~Dall'Agata, K.~Lechner, and M.~Tonin, ``{Covariant actions for N=1, D = 6
  supergravity theories with chiral bosons},''
  \href{http://dx.doi.org/10.1016/S0550-3213(97)00742-6}{{\em Nucl.Phys.}
  {\bfseries B512} (1998) 179--198},
\href{http://arxiv.org/abs/hep-th/9710127}{{\ttfamily arXiv:hep-th/9710127
  [hep-th]}}.

\bibitem{Dall'Agata:1998va}
G.~Dall'Agata, K.~Lechner, and M.~Tonin, ``{D = 10, N = IIB supergravity:
  Lorentz invariant actions and duality},'' {\em JHEP} {\bfseries 9807} (1998)
  017,
\href{http://arxiv.org/abs/hep-th/9806140}{{\ttfamily arXiv:hep-th/9806140
  [hep-th]}}.

\bibitem{DePol:2000re}
G.~De~Pol, H.~Singh, and M.~Tonin, ``{Action with manifest duality for
  maximally supersymmetric six-dimensional supergravity},''
  \href{http://dx.doi.org/10.1016/S0217-751X(00)00182-6}{{\em Int.J.Mod.Phys.}
  {\bfseries A15} (2000) 4447--4462},
\href{http://arxiv.org/abs/hep-th/0003106}{{\ttfamily arXiv:hep-th/0003106
  [hep-th]}}.

\end{thebibliography}

\providecommand{\href}[2]{#2}\begingroup\raggedright\endgroup

\end{document}